\documentclass{aa}
\usepackage[varg]{txfonts}
\usepackage{graphicx}
\usepackage{amsmath}
\usepackage{natbib}
\begin{document}

\title{Probing X-ray emission in different modes of PSR J1023+0038 with a radio pulsar scenario}
\author{S. Campana\inst{1} \and 
A. Miraval Zanon\inst{1,2} \and 
F. Coti Zelati\inst{3,4}  \and 
D. F. Torres\inst{3,4,5}  \and 
M. C. Baglio\inst{6,1} \and
A. Papitto\inst{7} }
\institute{
INAF-Osservatorio astronomico di Brera, Via Bianchi 46, I-23807, Merate (LC), Italy \\ \email{sergio.campana@brera.inaf.it}
\and
Universit\`a dell'Insubria, Dipartimento di Scienza e Alta Tecnologia, Via Valleggio 11, I-22100 Como, Italy
\and
Institute of Space Sciences (ICE, CSIC), Campus UAB, Carrer de Can Magrans, E-08193, Barcelona, Spain
\and 
Institut d'Estudis Espacials de Catalunya (IEEC), Gran Capit\`a 2-4, E-08034, Barcelona, Spain
\and
Instituci\'o Catalana de Recerca i Estudis Avan\c cats (ICREA), 08010 Barcelona, Spain
\and
New York University of Abu Dhabi, P.O. Box 129188, Abu Dhabi, UAE
\and
INAF-Osservatorio Astronomico di Roma, via Frascati 33, I-00040, Monteporzio Catone (Roma), Italy}
\date{May 2019}

\abstract{
Transitional pulsars provide us with a unique laboratory to study the physics of accretion onto a magnetic neutron star. PSR J1023+0038 (J1023) is the best studied of this class. We investigate  the X-ray spectral properties of J1023 in the framework of a working radio pulsar during the active state. We modelled the X-ray spectra in three modes (low, high, and flare) as well as in quiescence, to constrain the emission mechanism and source parameters. The emission model, formed by an assumed pulsar emission (thermal and magnetospheric) plus a shock component, can account for the data only adding a hot dense absorber covering \textasciitilde$30\%$ of the emitting source in high mode. The covering fraction is similar in flaring mode, thus excluding total enshrouding, and decreases in the low mode despite large uncertainties. This provides support to the recently advanced idea of a mini-pulsar wind nebula (PWN), where X-ray and optical pulsations arise via synchrotron shock emission in a very close ($\sim 100$  km, comparable to a light cylinder), PWN-like region that is associated with this hot absorber. In low mode, this region may expand, pulsations become undetectable, and the covering fraction decreases.
}

\titlerunning{An active radio pulsar for PSR J1023+0038 modes}
\authorrunning{S. Campana et al.}

\maketitle

\section{Introduction}

Transitional pulsars are a new class of variable neutron star X-ray binaries that alternate between periods of quiescence during which a radio pulsar is detected, to a more active state during which the source is brighter in X-rays and $\gamma$-rays, to a period in which disc-like features appear in the optical band (see Campana \& Di Salvo 2018 for a review).

During the active state, a puzzling behaviour has been observed across the multi-wavelength spectrum: PSR J1023+0038 (J1023 in the following) and the other transitional pulsars alternate between a high mode, during which X-ray pulsations are detected (with a pulsed fraction of $\sim8\%$) to a low mode, during which X-ray pulsations are not detected ($3\sigma$ upper limit of $<2.7\%$) and the source is dimmer in X-rays by a factor of approximately seven (Archibald et al. 2009, 2010, 2015; Papitto et al. 2015; Bogdanov et al. 2015; Campana et al. 2016).
%
Transitions from high to low mode occur very quickly on a $\sim 10$ s timescale in X-rays (Bogdanov et al. 2015). 
Similar rectangular dips occur in the optical and UV light curves (Shahbaz et al. 2015; Papitto et al. 2019; Jaodand et al. 2019).
A radio flare is often observed at the egress of low mode (Bogdanov et al. 2018). 
%
A flaring mode also appears during the active state; bright flares occur randomly and reach a luminosity a factor of approximately ten larger than in high mode with no detectable X-ray pulsations ($<1.5\%$; Archibald et al. 2015).

The detection of X-ray pulsations during the active state, has led several authors to speculate that these pulsations were caused by accretion of matter onto the neutron star surface 
(Linares 2014; Takata et al. 2014; Archibald et al. 2015; Bogdanov et al. 2015; Papitto et al. 2015; Campana et al. 2016; Jaodand et al. 2016; Coti Zelati et al. 2018).
%
Mechanisms involving the action of the neutron star rotating magnetosphere (i.e. propeller) were invoked to explain this astonishing behaviour along with the speed of the transition across modes (Papitto et al. 2014; Linares 2014; Papitto \& Torres 2015; Campana et al. 2016).

Observations with the Silicon Fast Astronomical Photometer (SiFAP) ultra-fast photometer recently revealed optical pulsations in J1023. 
Optical pulsations were detected at a low level ($\sim 1\%$ pulsed fraction) in high mode and at a lower amplitude in flaring mode, but not in low mode (Ambrosino et al. 2017; Papitto et al. 2019, see also Zampieri et al. 2019 who used a different fast photometer, Aqueye+).
 The detection of optical pulsations during the high mode of J1023 at such a high flux level is left unexplained if based on accretion 
(Ambrosino et al. 2017).
%
The most promising explanation for the optical pulsations in the accretion scenario should involve cyclotron emission in the accretion column. However, this mechanism fails to account for the observed pulsed luminosity by a factor of $>40$ (Papitto et al. 2019).
%
Pulsed emission has also been detected in the hard X-ray energy range with the Nuclear Spectroscopic Telescope Array ({\it NuSTAR}; Papitto et al. 2019)
and in the UV band with 
the Hubble Space Telescope ({\it HST}; Jaodand et al. 2019).
The spectrum of the pulsed emission can be described by a single power law from optical to hard X-rays with the same overall spectrum as the primary emission. 

Papitto et al. (2019) 
suggested that optical and X-ray pulses during high mode are produced by synchrotron emission from the intra-binary shock that forms where a striped pulsar wind meets the accretion disc, just outside the light cylinder. 
During low mode, this intra-binary shock moves away from the light cylinder so that the deposited energy is not enough to generate pulsations (but see Veledina et al. 2019, 
for a slightly different interpretation).
%
This new framework comprises a neutron star loosing energy through dipole losses during the active state, with the accreting matter outside the light cylinder (i.e. with all the conditions for a radio pulsar to be at work). In this case, pulsations do not come from the accretion of matter on the neutron star surface, rather they are ultimately amenable to the rotational energy losses (see also Coti Zelati et al. 2014).
%
Papitto et al. (2019) suggested that the flaring mode could be ascribed to a complete enshrouding of the radio pulsar, too.

In this Letter, we reconsider our accretion scenario for J1023 (Campana et al. 2016)
in light of these new discoveries. 
Based on X-ray data collected with the {\it XMM-Newton} satellite, we  test on the possible continuous presence of an active radio pulsar. We also add flaring mode data that were previously neglected.

\section{{\it XMM-Newton} data analysis}

We started from the same data set presented in Campana et al. (2016).
%
To these spectra obtained for J1023 in the low and high modes
(six {\it XMM-Newton} observations with the pn in timing mode
and the MOS in small
window mode), as well as
in quiescence (one {\it XMM-Newton} observation
and one {\it Chandra} observation), we added the flaring mode data of the six {\it XMM-Newton}
observations.  

The low energy part of the X-ray spectrum is of crucial importance in the spectral modelling (see below). 
For this reason, we considered only MOS data in the spectral fitting. 
Indeed, pn data in timing mode are trustworthy only starting from 0.6 keV, thus biasing the spectral fit towards the continuum and disadvantaging the absorption component.

In brief, data were grade-filtered using pattern 0-12 for MOS data and FLAG==0 and \#XMMEA\_EM options.
Proton flares were filtered out using the standard recipe. The source events were extracted using an 870 pixel circular region.
Following Bogdanov et al. (2015), good time intervals were generated for low mode for a source count rate (pn+2 MOS) in the 0.0-2.1 c s$^{-1}$ interval, for high mode in the 4.1-11 c s$^{-1}$, and for flaring mode in $>15$ c s$^{-1}$. The respective ancillary and redistribution files were then generated for each spectrum. 
The MOS spectra (0.3-10 keV) were rebinned to 100 counts per energy bin; MOS data comprise about 0.5 million of photons. Given the very large signal to noise of our spectra, we added a systematic error of $2\%$ to each spectral bin. 

\begin{figure*}
    \includegraphics[width=0.8\textwidth,angle=-90]{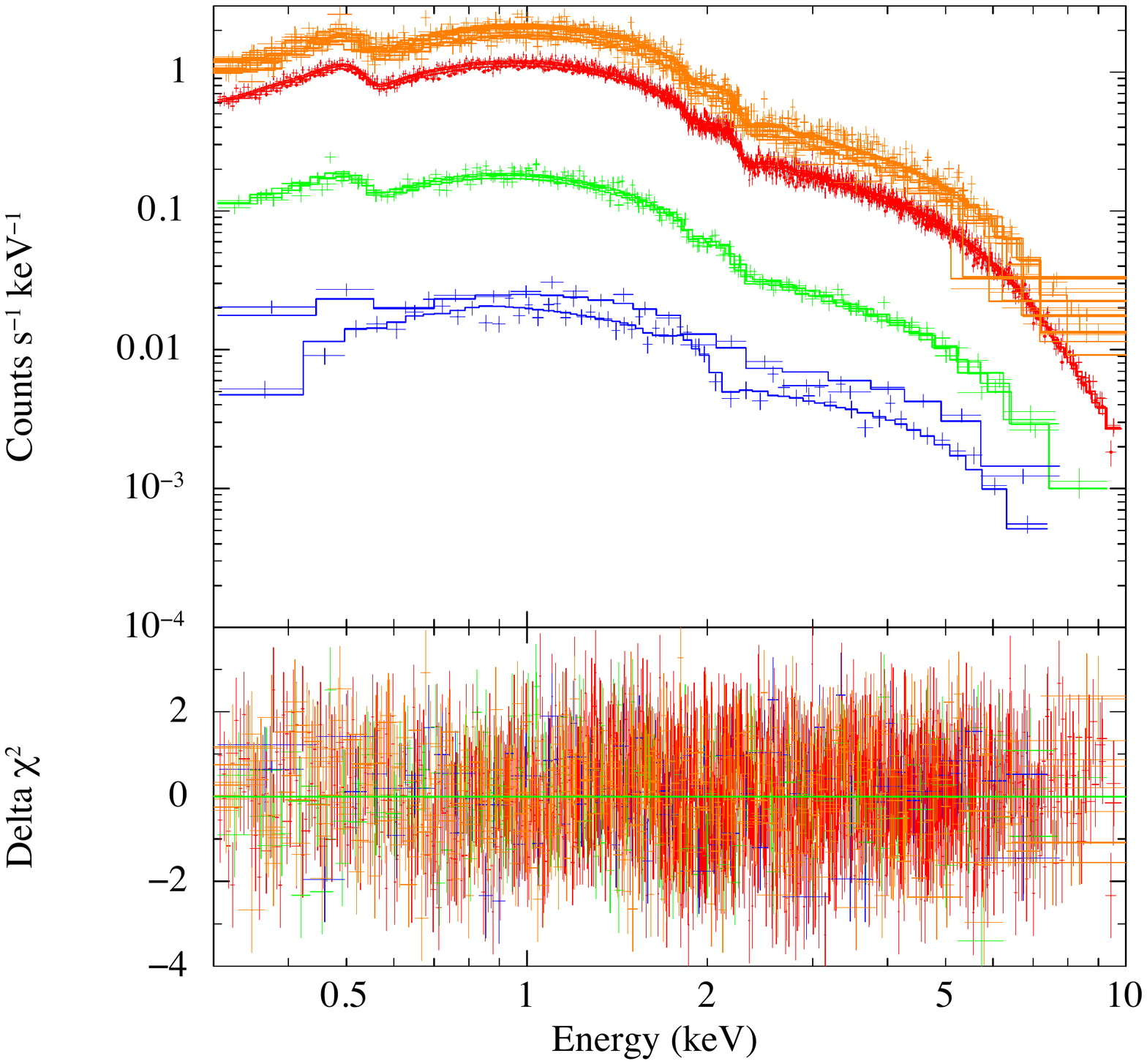}
    \caption{{\it XMM-Newton} spectra of J1023. Spectra are shown in the top panel. Orange refers to flare data, red to high mode data, green to low mode data, and blue (one {\it XMM-Newton} and one {Chandra}) to data in quiescence. In the bottom panel residuals are shown with the same colour coding.}
    \label{fig:my_label}
\end{figure*}

\section{Physical scenario}

The physical scenario driven by the new optical observations involves an active pulsar during the active state in all modes. 
Building the overall spectrum, the spectral component that we assume is present in all the spectra (quiescent, low mode, high mode, and flaring mode) is that associated with the pulsar. 
We model this as the contribution from a hot spot, i.e. a neutron star atmosphere model with free radius (NSATMOS; Heinke et al. 2006),
and a magnetosphere (power law; PL1).
If a pulsar is active, its relativistic particle wind interacts with the surrounding medium to produce high energy photons in a shock front (shock emission). 
This has been deeply investigated in the context of radio pulsars in binary systems (Arons \& Tavani 1993; Tavani \& Arons 1997; Takata et al. 2014, 2017).
%
This shock emission, which is modelled with a power-law component, depends on the geometry and density of the shock front. 
Therefore, we need a different shock emission component for the high and low mode spectra. 
This component is largely dominant in the high mode spectrum, but in the low mode spectrum both the shock and pulsar components are comparable. 
The power law is fixed in shape and normalisation for each low mode observation and for each high mode observation. Finally, to account for flare emission we added a different power-law component with the same photon index, but different normalisation to account for the different flare levels.
These prescriptions set the source spectrum in all four modes: quiescent, low, high, and flare.
This emission is observed to be absorbed by the interstellar medium; the absorption model {\tt tbabs} is used to model it equally for all the modes.

Fitting all the different mode data together with this overall model returns a very poor fit; i.e. a reduced $\chi^2=1.18$ for 2619 degrees of freedom, corresponding to a null hypothesis probability of $10^{-10}$. 
Wavy residuals in the spectral fit are present with a downturn at low energies. 
These residuals might be ascribed to the matter close to the light cylinder, which causes a distinctive absorption pattern. 
To account for this, we included the photo-ionised absorption model {\tt zxipcf}. This component should be absent during quiescence and may vary among the three modes.
While the column density (i.e. the amount of matter along the line of sight) should be the same, the ionisation parameter $\xi=L/nr^2$, where $L$ is the ionising luminosity, $n$ the density of the medium, and $r$ the distance of the medium, and the covering factor $f$, could be different. We expect the same column density because the ionisation state or distance from the source can change, but the total amount cannot change on a $\sim 10$ s timescale.  
We would a priori expect a larger ionisation and covering factor in high mode with respect to the low mode values. 
This inference comes from Papitto et al. (2019),
where they suggested that in high mode matter is close to the light cylinder, whereas during low mode episodes matter is pushed further out, preventing the pulsations from occurring.
%
Papitto et al. (2019) proposed that during the flaring mode events the source would be completely enshrouded, which must result in a large covering factor.

\begin{table*}[!ht]
    \caption{Spectral fit parameters.}
    \centering
    \begin{tabular}{c|c|c|c|c}
\hline
Parameter & Flaring mode  & High mode & Low mode & Quiescence\\
\hline
Column density ($10^{20}$ cm$^{-2}$) & $5.0^{+0.3}_{-0.2}$ & tied & tied & tied\\
Col. dens. Hot absorber ($10^{22}$ cm$^{-2}$) & $23.4^{+5.7}_{-1.0}$ & tied & tied & 0 \\
Covering fraction ($f_{\rm cov}$)  &$0.27^{+0.07}_{-0.05}$ & $0.27^{+0.04}_{-0.01}$& $0.18^{+0.15}_{-0.06}$& 0 \\
Ionisation par. ($\log{\xi}$)      &$1.4^{+0.8}_{-1.5}$ & $1.9^{+0.1}_{-0.2}$& $1.9^{+0.2}_{-2.0}$& 0 \\
Shock emission photon index & $1.76^{+0.02}_{-0.02}$ & 
$1.82^{+0.01}_{-0.01}$ &  
$1.99^{+0.02}_{-0.03}$ & 
-- \\
Thermal component $\log(T/K)$ & tied & tied & tied & 
$5.85^{+0.13}_{-0.10}$\\ 
Emitting fraction $f$ & tied & tied & tied & $0.16^{+0.18}_{-0.07}$\\
Magnetospheric emission ($\Gamma_{\rm mag}$) & tied & tied & tied & $1.06^{+0.06}_{-0.08}$\\
\hline
Absorbed flux ($10^{-12}$ erg cm$^{-2}$ s$^{-1}$) 0.3-10 keV & 
50.2 & 13.6 & 1.99 & 0.54\\ 
Unabsorbed flux ($10^{-12}$ erg cm$^{-2}$ s$^{-1}$) 0.3-10 keV & 
69.2 & 18.6 & 2.54 & 0.56\\
Luminosity  ($10^{32}$ erg s$^{-1}$) 0.3-10 keV & 
156 & 42.2 & 5.76 & 1.27\\
\hline
    \end{tabular}
    \label{tab:spectral}
    
Errors are calculated with $\Delta\chi^2=2.70$. The fit provided a $\chi^2_{\rm red}=1.07$ for 2612 degrees of freedom. \\
Fluxes for the flaring mode refer to a mean of the 5 spectra (with a very small standard deviation of $4\%$). Luminosity was calculated adopting a distance of 1.37 kpc (Deller et al. 2012).
\end{table*}

\section{Spectral fitting}

We fit the MOS12 data of quiescence, low, high, and flaring modes (plus one Chandra spectrum for the quiescence of J1023) with the spectral model {\tt tbabs $\times$ zxipcf $\times$ (nsatmos + powerlaw + powerlaw)}, as detailed in the previous section. An inter-calibration constant was also included.
The results are summarised in Table \ref{tab:spectral}.

Fig. 1 shows the result of our fitting procedure. The overall fit is relatively good with a reduced $\chi^2=1.07$ for 2612 degrees of freedom corresponding to a formal $0.5\%$ null hypothesis probability. 
We remark, however, that these data were collected across several years so that small changes in the spectral parameters could be envisaged, degrading the goodness of the fit.
For a neutron star mass of $1.4\,M_\odot$ and radius 10 km, the hot emission region has a temperature of $\log(T/K)=5.85^{+0.13}_{-0.10}$ (obtained for $\Delta\chi^2=2.71$) and fraction of the emitting surface of $f=0.16^{+0.18}_{-0.07}$. The magnetospheric power law is hard as often observed in radio pulsars with a photon index of $\Gamma_{\rm mag}=1.06^{+0.06}_{-0.08}$. The 0.3-10 keV unabsorbed quiescent luminosity is $1.3\times 10^{32}$ erg s$^{-1}$, adopting a distance of 1.37 kpc 
(Deller et al. 2012), the non-thermal component comprises $92\%$ of the total (see Table \ref{tab:spectral}). 

The high mode has, in addition, a power-law component with photon index $\Gamma_{\rm high}=1.82\pm0.01$. The 0.3-10 keV high mode luminosity is $4.2\times 10^{33}$ erg s$^{-1}$ with the power law making $97\%$ of the total. In the low mode, the power-law component flattens to $\Gamma_{\rm low}=1.99^{+0.02}_{-0.03}$. 
The 0.3-10 keV low mode luminosity is $5.8\times 10^{32}$ erg s$^{-1}$; the power law comprises $22\%$ of the total. The flaring mode reaches a higher (mean) luminosity $1.6\times 10^{34}$ erg s$^{-1}$; the power-law component with $\Gamma_{\rm flare}=1.76^{+0.02}_{-0.02}$ comprises the entire flux (Table \ref{tab:flux}).

Interesting results are obtained for the absorption components. The interstellar absorption is $5.0^{+0.3}_{-0.2}\times 10^{20}$ cm$^{-2}$, which is consistent with tabulated values 
(Willingale et al. 2003).
%
The intra-binary absorption component is set to zero for quiescence (and if left free to vary is $<0.5$, $90\%$ confidence level). 
When J1023 is active, the hot absorber has a large column density of $2.34^{+0.59}_{-0.09}\times 10^{23}$ cm$^{-2}$. This is tied to all the modes. 
In low mode the ionisation parameter is $\log{\xi}=1.9^{+0.2}_{-2.0}$ with a large error. The covering factor of this hot absorber is $0.18^{+0.15}_{-0.06}$. In contrast to high mode, the 
ionisation parameter is comparable to $1.9^{+0.1}_{-0.2}$ (despite a much smaller error), but the covering factor, $f_{\rm cov}=0.27^{+0.04}_{-0.01}$, is larger.
This should indicate an absorber closer to the source in high mode and further away in low mode.
Interestingly, during flaring mode, the ionisation parameter drops somewhat $1.4^{+0.8}_{-1.5}$ (still within the errors), and the covering factor remains consistent with high mode value, $0.27^{+0.07}_{-0.05}$. 
This indicates that during flaring mode the source is not fully enshrouded by this hot medium. 
Instead, flares may be due to disc or companion star activity, possibly related to magnetic reconnection.

\begin{table}[h]
    \centering
    \caption{Percentage of the contribution of spectral components to the overall unabsorbed 0.3-10 keV luminosity.}
    \begin{tabular}{c|ccc}
\hline
State                           & NSA & Magn. Em. & PL \\
\hline
Flaring mode ($1.6\times 10^{34}$)& 0\% & 0\%       & 100\% \\
High mode  ($4.2\times 10^{33}$)& 0\% &  3\%      & 97\%  \\    
Low mode   ($5.8\times 10^{32}$)& 6\% & 72\%      & 22\% \\
Quiescence ($1.3\times 10^{32}$)& 8\% & 92\%      & -- \\
\hline
    \end{tabular}
    \label{tab:flux}
\end{table}

\section{Conclusions}

In this Letter, we have analysed the {\it XMM-Newton} and {\it Chandra} X-ray spectra of the transitional pulsar J1023. 
These spectra encompass all the states showed by J1023: the quiescent and active, conjugated in three flavours, low, high, and flaring mode. 
The rationale guiding our fit was the recently discovered optical pulsations (Ambrosino et al. 2017; Papitto et al. 2019).
%
These pulsations are present during high mode only and are not neatly explained in terms of accretion of matter onto the neutron star. 
Accretion models fail by at least an order of magnitude to account for the observed pulsed luminosity (Ambrosino et al. 2017).
%
A model in which X-ray pulsations occur during high mode (only) has been put forwards by Papitto et al. (2019).
%
This model involves a pulsar that is always at work during the active state of J1023 and ultimately relates the pulsation to the rotational power.
%

Our X-ray spectral analysis tries to account for the observed spectra in the radio pulsar scenario. 
It is by no means unique (e.g. see Campana et al. 2016)
for a different working spectral model), but it aims at testing whether a pulsar can be at work at all times.
We find that the spectra can be modelled with a typical pulsar spectrum  that has a thermal component plus a magnetospheric component, which is needed to account for the quiescent and low mode spectra, 
plus a power-law component, which dominates the high and flaring mode spectra but is also needed for modelling the low mode spectra.

This power-law component can be interpreted in terms of shock emission between the relativistic pulsar wind and the incoming material in high and low modes.
As expected it is steeper ($\Gamma\sim 2$) at lower luminosity (low mode) and flatter at higher luminosity ($\Gamma\sim 1.8$), as observed in pulsars interacting with massive companions, as PSR B1259+63 (Tavani \& Arons 1997).
%
The shock front in low mode lies at larger distances from the light cylinder region, the geometry for generating pulsations is not favourable, and a classical shock is expected. In high mode, instead, the front gets close to the light cylinder, pulsations may be generated according to the mechanism proposed by Papitto et al. (2019), 
and the shock emission gets harder. 
The fact that the shock component is by far dominant in high mode supports this mini-pulsar wind nebula (PWN) interpretation advanced by Papitto et al. (2019).

In this scenario, the pulsations are not magnetospheric, but are still related to the rotation power.
Pulses arise via synchrotron emission in a very close, mini-PWN-like region of the dimension of the light cylinder ($\sim 100$ km). 
Optical pulsations come from this mechanism too, thus alleviating the problem of high conversion efficiency into the optical band.
As J1023 changes to low mode, the PWN expands and pulsations vanish (Papitto et al. 2019).
Our scenario is different from that proposed by Veledina et al. (2019),
who suggested that the low mode might be associated with the propeller. 
A weak indication in favour of the PWN expansion in low mode comes from simulations (Parfrey \& Tchekhovskoy 2017), 
showing that in the propeller 
state the covering angle of the surrounding medium should increase. Instead, we found a (low significance) decrease in the covering fraction.

In the optical, there is evidence for the presence of an accretion disc at large distances from the neutron star (Hernandez Santisteban 2016).
%
This disc is truncated and evaporated by the pulsar pressure just outside the light cylinder in high mode. Solutions for this kind of disc geometry exist (Ek\c{s}i \& Alpar 2005).
In low mode, the disc is pushed further out to recollapse again at transition.
In agreement with this, the covering factor of this hot ambient medium related to the accretion disc intercepts a larger fraction of the emitting region in high mode (being closer) than in low mode (being more distant). 

There have been suggestions that the flaring mode might correspond to a complete enshrouding of the pulsar. However, even if the power-law index is flatter in flaring mode, the covering fraction of the hot surrounding medium remains at a level comparable to that of high mode. We suggest instead magnetic reconnection in the disc or even in the disc/star as a possible cause of flaring activity. 
Indeed, pulsations were not detected in X-ray during flaring mode, but they were in the optical. The pulsed fraction is lower than in high mode by a factor of approximately five, but similar in shape, whereas the X-ray flux increases by a factor of  approximately four, suggesting that dilution might be an explanation. 
We investigate the flaring mode of another transitional pulsar XSS J12270--4859 in more detail, which occurs with a soft and hard spectrum (de Martino et al. 2015; Miraval Zanon et al. 2019).

The most intriguing characteristic of the proposed spectral model is the presence of a hot medium. Indirect evidence for the presence of this medium comes from the detection of narrow emission and absorption lines in the RGS spectra of J1023. These lines were described in Coti Zelati et al. (2018),
even though it was not possible to search for differences between the low and high modes owing to limited statistics. The N VI triplet allowed us to set a lower limit on the particle density of $n>10^{11}$ cm$^{-3}$, suggesting a dense medium.
Clearly, our identification of a unique medium with a single density (and ionisation parameter) is an oversimplification of a medium that could be more complex and stratified in latitude and turbulent. Despite these caveats, our spectral fitting provides a supportive indication to the proposed pulsar/mini-PWN scenario.

\subsection*{Acknowledgments}
We thank the referee for a careful reading of the manuscript and for useful comments.
DFT and FCZ acknowledge support from grants PGC2018-095512-B-I00, SGR2017-1383, and AYA2017-92402-EXP. AP acknowledges  financial support from the Italian Space Agency and National Institute for Astrophysics, ASI/INAF, under agreements ASI-INAF I/037/12/0 and ASI-INAF 2017-14-H.0.


\begin{thebibliography}{}

\bibitem[Ambrosino, Papitto et al. (2017)] {ambrosino17}
Ambrosino, F., Papitto, A., Stella, L., et al. 2017, Nat. Astr. 1, 854 

\bibitem[Archibald et al. (2009)] {archibald09} 
Archibald, A. M., Stairs, I. H., Ransom, S. M., et al. 2009, Science, 324, 1411 

\bibitem[Archibald et al. (2010)] {archibald10} 
Archibald, A. M., Kaspi, V. Bogdanov, S.,  Hessels, J. W. T., Stairs, I. H., Ransom, S. M., McLaughlin, M. A. 2010, ApJ, 722, 88

\bibitem[Archibald et al. (2015)] {archibald15} 
Archibald, A. M., Bogdanov, S., Patruno, A., et al. 2015, ApJ, 807, 62 

\bibitem[Arons \& Tavani (1993)] {arons93} 
Arons, J., \& Tavani, M. 1993, ApJ, 403, 249 

\bibitem[Bogdanov et al. (2015)] {bogdanov15} 
Bogdanov, S., Archibald, A. M., Bassa, C., et al. 2015,  ApJ, 806, 148

\bibitem[Bogdanov et al. (2018)] {bogdanov18} 
Bogdanov, S., Deller, A. T., Miller-Jones, J. C. A., et al. 2018, ApJ, 856, 54

\bibitem[Campana et al. (2016)] {campana16} 
Campana, S., Coti Zelati, F., Papitto, A., et al. 2016, A\&A, 594, A31

\bibitem[Campana \& Di Salvo (2018)] {campana18} 
Campana, S., \& Di Salvo, T. 2018, ArXiv e-prints, arXiv:1804.03422 

\bibitem[Coti Zelati et al. (2018)] {coti18} 
Coti Zelati, F., Campana, S., Braito, V., et al. 2018, A\&A., 611, A14

\bibitem[Coti Zelati et al. (2014)] {coti14} 
Coti Zelati, F., Baglio, M. C., Campana, S., et al. 2014,  MNRAS, 444, 1783

\bibitem[De Martino et al. (2015)] {demartino15} 
de Martino, D., Papitto, A., Belloni, T., 
et al. 2015,  MNRAS, 454, 2190

\bibitem[Deller et al. (2012)] {deller12} 
Deller, A. T., Archibald, A. M., Brisken, W. F., et al. 2012, ApJ, 756, L25

\bibitem[Ek\c{s}i \& Alpar (2005)] {eksi05} 
Ek\c{s}i, K. Y., \& Alpar, M. A. 2005, ApJ 620, 390

\bibitem[Hernandez Santisteban (2016)] {hernandez16} 
Hernandez Santisteban, J. V. 2016, PhD Thesis Dissertation (U. Southampton)

\bibitem[Heinke et al. (2006)] {heinke06} 
Heinke, C. O., Rybicki, G. B., Narayan, R., Grindlay, J. E. 2006, ApJ, 644, 1090

\bibitem[Jaodand et al. (2016)] {jaodand16} 
Jaodand, A., Archibald, A. M., Hessels, J. W. T., et al. 2016, ApJ, 830, 122

\bibitem[Jaodand et al. (2019)] {jaodand19} 
Jaodand, A., et al. 2019, ApJ, submitted

\bibitem[Linares (2014)] {linares14} 
Linares, M. 2014, ApJ, 795, 72

\bibitem[Miraval Zanon et al. (2019)] {miraval19} 
Miraval Zanon, A., Campana, S., Ridolfi, A. 2019, A\&A, submitted

\bibitem[Papitto et al. (2014)] {papitto14} 
Papitto, A., Torres, D. F., Li, J. 2014,  MNRAS, 438, 2105

\bibitem[Papitto et al. (2015)] {papitto15} 
Papitto, A., de Martino, D., Belloni, T. M., et al. 2015,  MNRAS, 449, L26

\bibitem[Papitto \& Torres (2015)] {papitto_torres15} 
Papitto, A., Torres, D. F. 2015, ApJ 807, 33

\bibitem[Papitto et al. (2019)] {papitto19} 
Papitto, A., Ambrosino, F., Stella, L., et al. 2019, ApJ in press (arXiv:1904.10433)

\bibitem[Parfrey \& Tchekhovskoy (2017)] {parfrey17} 
Parfrey, K., \& Tchekhovskoy, A. 2017, ApJ, 851, L34


\bibitem[Shahbaz et al. (2015)] {shahbaz15} 
Shahbaz, T., Linares, M., Nevado, S. P., et al. 2015, MNRAS, 453, 3461

\bibitem[Takata et al. (2014)] {takata14} 
Takata, J., Li, K. L., Leung, G. C. K., et al. 2014, ApJ, 785, 131 

\bibitem[Takata et al. (2017)] {takata17} 
Takata, J., Tam, P. H. T., Ng, C. W. 2017, ApJ, 836, 241

\bibitem[Tavani \& Arons (1997)] {tavani97} 
Tavani, M., \& Arons, J. 1997, ApJ 477, 439

\bibitem[Veledina et al. (2019)] {veledina19} 
Veledina, A., N\"attil\"a, J., Beloborodov, A. M. 2019, ApJ submitted (arXiv:1906.02519)

\bibitem[Willingale et al. (2003)] {willingale03} 
Willingale, R., Starling, R. L. C., Beardmore, A. P., Tanvir, N. R., O'Brien, P. T. 2003, MNRAS, 431, 394

\bibitem[Zampieri et al. (2019)] {zampieri19} 
Zampieri, L., Burtovoi, A., Fiori, M., et al. 2019, MNRAS 485, L109

\end{thebibliography}
\end{document}